\documentclass[preprint,noshowkeys,prl,showpacs]{revtex4}
\usepackage{amssymb}
\usepackage{amsmath}
\usepackage{graphicx}
\usepackage{dcolumn}
\usepackage{color}

\begin{document}

\title{Ultrasensitive Detection of Majorana Fermions via Spin-based
Optomechanics with Carbon Nanotubes}
\author{Hua-Jun Chen}
\author{Ka-Di Zhu}
\email{zhukadi@sjtu.edu.cn}
\affiliation{Key Laboratory of Artificial Structures and Quantum Control (Ministry of
Education), Department of Physics, Shanghai Jiao Tong University, 800
DongChuan Road, Shanghai 200240, China }
\date{\today }

\begin{abstract}
We propose a novel optical method to detect the existence of Majorana
fermions at the ends of the semiconductor nanowire via the coupling to an
electron spin trapped on a carbon nanotube resonator under the control of a
strong pump field and a weak probe field. The coupling strength of Majorana
fermion to the spin in the carbon nanotube and the decay rate of the
Majorana fermion can be easily measured from the probe absorption spectrum
via manipulating the spin-mechanical coupling in the suspended\ carbon
nanotube. The scheme proposed here will open a good perspective for its
applications in all-optical controlled Majorana fermion-based quantum
computation and quantum information processing.
\end{abstract}

\maketitle




\section{INTRODUCTION}

Majorana fermions (MFs) proposed by Majorana\cite{Majorana} in 1937, with
zero energy quantum particles described by hermitian second quantized
operators $\gamma ^{\dagger }=\gamma $, have attracted a great deal of one's
attention due to their potential applications, e.g., topological quantum
computation\cite{Alicea0,Beenakker,Tudor,Franz}. MFs, although originally
proposed as a model for neutrinos, which have been detected in condensed
matter systems\cite{Sarma,Stern,Bonderson,Lutchyn,Refael,Sau0}. It has been
demonstrated that Majorana particles could exist at the ends of 1D nanowires
in the presence of the appropriate superconducting pairing\cite{Refael,Sau0}
under suitable conditions. Not long ago, some researches show that using a
heterostructure consisted of two conventional materials such as s-wave
superconductor and semiconductor nanowire with strong spin-orbit coupling,
one can generate MFs \cite{Sau1,Sau2,Alicea,Potter:prl10,Mao:prl12}.
Although a few schemes have been proved that MFs dwell in solid state
systems, experimental observation Majorana bound states (MBSs) are still a
challenging, and if the direct experimental observation of Majorana
particles is achieved\ in solid state systems, there would be a breakthrough
on the research of fundamental physics and the future quantum computer\cite%
{Tudor}. Fortunately, very recently, some experiment reports have been shown
that the existence of a pair of MBSs are observed at the ends of the
different materials nanowire\ which contact with an superconductor via zero
bias conductance peak\cite{Deng,Mourik:sci12,Das} and fractional AC
Josephson effect\cite{Rokhinson}.\ Actually, besides these schemes, MBSs
could also be probed via quantum non-locality, transconductance, interference%
\cite{Tudor}, and the noise measurements\cite{Bolech,Nilsson,Law}, as well
as 4$\pi $ periodic Majorana-Josephson currents\cite{Fu}.

Up to now, the detection of MFs in condensed matter systems based on the
electrical measurements is the mainstream approach\cite%
{Deng,Mourik:sci12,Das,Rokhinson}. However, how to detect and verify the
definite MBSs is still challenging\cite{Franz}, because the noises that may
confuse the MBSs signals\ will be involved in the electrical measurement
schemes, and they will affect the judgement whether\ the Majorana-like
signals are true MBSs. Moreover, these methods also include the electron
transfer into or out of MBSs via the nanowire, which will destroy the qubit
information when MBSs are used to quantum computing. In this case, one
proposes an experimental setup for detecting a Majorana zero mode by
employing a single quantum dot (QD) coupled to the end of a p-wave
superconducting nanowire\cite{Liu:prb11,Cao:prb12}. This scheme provides a
potential way to probe MBSs without totally destroying the information in
the qubit. Besides, Leijnse et al. present\ a setup to measure the lifetime
of Majorana bound states via a QD coupled to a semiconducting wire with a
superconductivity\cite{Leijnse}.

On the other hand, the suspended carbon nanotubes (CNTs), due to their
remarkable electronic properties, low masses and high-Q factors \cite%
{Witkamp:nl06,Jespersen:NP11,Eichler:NN11}, attract more and more people's
attention. These advantages will make the CNTs more applicable to
ultrasensitive detection and mechanical signal processing \cite%
{Benjamin:sci09}. Recently, the investigation of spin-orbit interaction
(SOI) in CNTs has gained significant advances, which pave the way for
manipulating the spin degree of freedom \cite%
{Kuemmeth:nat08,Arcizet:NP11,Kuemmeth:MT10,Struck:prl12,Li:sr12}. Based on
these advantages of CNTs, rather than the detection scheme\cite%
{Mourik:sci12,Rokhinson,Das,Benjamin,Walter} and what Stanescu et al.\cite%
{Tudor} elaborated, here, we shall propose a novel optical measurement
scheme to detect the existence of Majorana fermion by using an electron spin
trapped on a carbon nanotube resonator as a sensitive probe. We employ the
optical pump-probe scheme (see Fig. 1) which has been demonstrated
experimentally\cite{Weis:sci10,Teufel,Safavi-Naeini} to detect Majorana
fermion via the probe-absorption spectrum. The vibration of CNT (i.e. the
phonon cavity) enhances the probe spectrum, which makes the Majorana fermion
more sensitive to be detected. Moreover, the scheme also provide a method to
measure the lifetimes of Majorana fermion distinguishing the electrical
measurement\cite{Leijnse}.

\section{THEORY}

The schematic setup is shown in Fig. 1, where an electron spin trapped on
the suspended CNT\cite{Kuemmeth:nat08} is employed to detect a pair of
Majorana fermions emerged\ at the ends of the tunnel-coupled nanowire\cite%
{Mourik:sci12}. In the presence of a magnetic field \textbf{B}, it
constitutes a well-defined two-level spin system (TLS) with an effective
spin-phonon coupling due to the interplay between SOI and flexural
vibrations of the nanotube\cite{Ohm,FlensbergK}. The vibrational modes of
nanotube resonator can be treated as phonon modes. The inset window of Fig.1
shows the two-level spin state, while dressing with the flexural modes of
CNT via the spin-phonon coupling. The spin-phonon coupling mechanism is
presented, where at long phonon wavelengths the deflection coupling is
dominated, while at short wavelengths the deformation potential coupling
should be dominated\cite{Rudner}.

For simplicity we consider only the deflection coupling mechanism, but note
that the approach can readily be extended to include both effects. The
Hamiltonian describing this system is \cite{FlensbergK,Rudner}%
\begin{equation}
H_{CNT}=\frac{\Delta _{so}}{2}\tau _{3}(\mathbf{S}\cdot \mathbf{t})+\Delta
_{KK^{^{\prime }}}\tau _{1}-\mu _{orb}\tau _{3}(\mathbf{B}\cdot \mathbf{t}%
)+\mu _{B}(\mathbf{S}\cdot \mathbf{B})\text{,}
\end{equation}%
where $\Delta _{so}$ and $\Delta _{KK^{^{\prime }}}$ represent the curvature
enhanced spin-orbit coupling and the inter-valley coupling, respectively. $%
\tau _{i}$ and $S$ are the Pauli matrices in valley and spin space, $\mathbf{%
t}$ is the tangent vector along the CNT axis, and $\mathbf{B=}Be_{z}$ is the
magnetic field along the axis of the suspended CNT. In order to study how
electron spin couples to the quantized mechanical motion of the CNT, we
consider only a single polarization of flexural motion, assuming that the
two-fold degeneracy is broken (for example, driven by an external electric
field). A generic deformation of the CNT with deflection $u(z)$ makes the
tangent vector $t(z)$ coordinate-dependent. Expanding $t(z)$ for small
deflections, the coupling terms in Hamiltonian (1) can be rewritten as $%
\mathbf{S}\cdot \mathbf{t\simeq }S_{z}+(du/dz)S_{x}$ and $\mathbf{B}\cdot
\mathbf{t\simeq }B_{z}+(du/dz)B_{x}$. Expressing the deflection $u(z)$ in
terms of the creation and annihilation operators $a^{+}$ and $a$ for a
quantized flexural phonon mode $u(z)=f(z)l_{0}(a+a^{+})/\sqrt{2}$, where $%
f(z)$ and $l_{0}$ are the waveform and zero-point amplitude of the phonon.
Thus, we obtain the Hamiltonian of the suspended CNT system as $%
H_{CNT}=\hbar \omega _{S}S^{z}+\hbar \omega _{n}a^{+}a+\hbar
g(S^{+}a+S^{-}a^{+})$, where $\hbar \omega _{S}=\mu _{B}(B-\Delta _{so}/2\mu
_{B})$, $B$ denotes the magnetic field strength and $\mu _{B}$ is spin
magnetic moment. $\omega _{n}$ is the vibrational frequency of the nanotube
resonator. $g$ is the spin-phonon coupling strength in CNT\cite{Struck:prl12}%
.

For the nanowire structure, combining a strong Rashba spin-orbit interaction
and Zeeman splitting, the proximity-effect-induced s-wave superconductivity
in the nanowire can support electron-hole quasiparticle excitations of MBSs
at the ends of the nanowire\cite{Refael,Lutchyn,Sau1}. Suppose the electron
spin couples to $\gamma _{1}$, then the Hamiltonian that describes the MBSs%
\cite{Sau0,Cao:prb12,Liu:prb11} is $H_{1}=i\hbar \omega _{M}\gamma
_{1}\gamma _{2}/2+i\hbar \beta (S^{-}-S^{+})\gamma _{1}$. To detect the
existence of Majorana fermion, it is helpful to switch from the Majorana
representation to the regular fermion one through the exact transformation $%
\gamma _{1}=f^{\dag }+f$ and $\gamma _{2}$ $=i(f^{\dag }-f)$. $f$ and $%
f^{\dag }$ are the fermion annihilation and creation operators obeying the
anti-commutative relation $\left\{ f\text{, }f^{\dag }\right\} =1$.
Accordingly, $H_{1}$ can be rewritten as $H_{MBS}=\hbar \omega _{M}(f^{\dag
}f-1/2)+i\hbar \beta (S^{-}f^{\dag }-S^{+}f)$, where the first term gives
the energy of Majorana fermion at frequency $\omega _{M}$, the second term
describes the coupling between the right MBSs and the electron spin in CNT
with the coupling strength $\beta $. Applying a strong pump field and a weak
probe field to the spin of an electron simultaneously. Then the Hamiltonian
of the electron spin coupled to the pump field and probe field is given by%
\cite{Boyd} $H_{S-F}=-\mu E_{pu}(S^{+}e^{-i\omega _{pu}t}+S^{-}e^{i\omega
_{pu}t})-\mu E_{pr}(S^{+}e^{-i\omega _{pr}t}+S^{-}e^{i\omega _{pr}t})$,
where $\mu $ is the dipole moment of the electron, $\omega _{pu}$ and $%
\omega _{pr}$ are the frequency of the pump field and the probe field,
respectively and $E_{pu}$ ($E_{pr}$) is the slowly varying envelope of the
pump field (probe field). Therefore, we can obtain the total Hamiltonian of
the system in the presence of two optical excitation\cite{Boyd,Cao:prb12}.
In a rotating frame at the pump field frequency $\omega _{pu}$, the total
Hamiltonian of the system can be written by
\begin{align}
H& =\hbar \Delta _{pu}S^{z}+\hbar \Delta _{n}a^{+}a+\hbar \Delta _{M}f^{\dag
}f+\hbar g(S^{+}a+S^{-}a^{+})+i\hbar \beta (S^{-}f^{\dag }-S^{+}f)  \notag \\
& -\hbar \Omega _{pu}(S^{+}+S^{-})-\mu E_{pr}(S^{+}e^{-i\delta
t}+S^{-}e^{i\delta t})\text{,}
\end{align}%
where $\Delta _{pu}=\omega _{D}-\omega _{pu}$ is the detuning of the exciton
frequency and the pump frequency, $\Omega _{pu}=\mu E_{pu}/\hbar $ is the
Rabi frequency of the pump field, and $\delta =\omega _{pr}-\omega _{pu}$ is
the probe-pump detuning. $\Delta _{M}=\omega _{M}-\omega _{pu}$ is the
detuning of the Majorana fermion frequency and the pump frequency. Actually,
we have neglected the regular fermion like normal electrons in the nanowire
that interact with the QD in the above discuss. To describe the interaction
between the normal electrons and the QD, we use the tight binding
Hamiltonian of the whole wire as \cite{Mahan}: $H_{fs}=\hbar \omega
_{D}S^{z}+\hbar \sum\limits_{k}\omega _{k}c_{k}^{+}c_{k}+\hbar \lambda
\sum\limits_{k}(c_{k}^{+}S^{-}+S^{+}c_{k})$, where $c_{k}$ and $c_{k}^{\dag
} $ are the regular fermion annihilation and creation operators with energy $%
\omega _{k}$ and momentum $k$ obeying the anti-commutative relation $\left\{
c_{k}\text{, }c_{k}^{\dag }\right\} =1$, and $g$ is the coupling strength
between the normal electrons and the QD ( here for simplicity we have
neglected the $k$-dependence of $g$ as in Ref.\cite{Hewson}).

According to the Heisenberg equation of motion $i\hbar dO/dt=[O,H]$ and
introducing the corresponding damping and noise terms, we derive the quantum
Langevin equations as follows \cite{Gardiner,Walls}%
\begin{align}
\dot{S}^{z}& =-\Gamma _{1}(S_{z}+1/2)-ig(S^{+}a-S^{-}a^{\dag })-\beta
(S^{-}f^{\dag }+S^{+}f)+i\Omega _{pu}(S^{+}-S^{-})  \notag \\
& +(i\mu E_{pr}/\hbar )(S^{+}e^{-i\delta t}-S^{-}e^{i\delta t})\text{,}
\end{align}%
\begin{equation}
\dot{S}^{-}=-(i\Delta _{pu}+\Gamma _{2})S^{-}+2(\beta f+iga)S^{z}-2i\Omega
_{pu}S^{z}-2i\mu E_{pr}S^{z}e^{-i\delta t}/\hbar +\hat{F}_{in}(t)\text{,}
\end{equation}%
\begin{equation}
\dot{f}=-(i\Delta _{M}+\kappa _{M}/2)f+\beta S^{-}+\hat{\xi}_{M}(t)\text{,}
\end{equation}%
\begin{equation}
\dot{a}=-(i\Delta _{n}+\kappa /2)a-igS^{-}+\hat{\xi}(t)\text{,}
\end{equation}%
where $\Gamma _{1}$ and $\Gamma _{2}$ are the electron spontaneous emission
rate and dephasing rate, $\kappa $ is the decay rate of CNT and $\kappa _{M}$
is the decay rate of the Majorana fermion. $\hat{F}_{in}(t)$ is the $\delta $%
-correlated Langevin noise operator, which has zero mean $\left\langle \hat{F%
}_{in}(t)\right\rangle =0$ and obeys the correlation function $\left\langle
\hat{F}_{in}(t)\hat{F}_{in}^{\dagger }(t^{^{\prime }})\right\rangle
\thicksim \delta (t-t^{^{\prime }})$. The resonator mode of CNT is affected
by a Brownian stochastic force with zero mean value, and $\hat{\xi}(t)$ has
the correlation function $\left\langle \hat{\xi}^{+}(t)\hat{\xi}(t^{^{\prime
}})\right\rangle =\dfrac{\kappa }{\omega _{n}}\int \dfrac{d\omega }{2\pi }%
\omega e^{-i\omega (t-t^{^{\prime }})}[1+\coth (\dfrac{\hbar \omega }{%
2\kappa _{B}T})]$, where $k_{B}$ and $T$ are the Boltzmann constant and the
temperature of the reservoir of the coupled system. Majorana fermion has the
same correlation relation as resonator mode of CNT.

To go beyond weak coupling, the Heisenberg operator can be rewritten as the
sum of its steady-state mean value and a small fluctuation with zero mean
value: $S^{z}=S_{0}^{z}+\delta S^{z}$, $S^{-}=S_{0}^{-}+\delta S^{-}$, $%
f=f_{0}+\delta f$ and $a=a_{0}+\delta a$. Since the driving fields are weak,
but classical coherent fields, we will identify all operators with their
expectation values, and drop the quantum and thermal noise terms\cite%
{Walls,Weis:sci10}. Simultaneously inserting these operators into the
Langevin equations Eqs. (3)-(6) and neglecting the nonlinear term $\delta
f\delta S^{-}$, $\delta a\delta S^{-}$, $\delta S^{z}\delta f$ and $\delta
S^{z}\delta a$, the linearized Langevin equations can be written as:%
\begin{align}
\left\langle \delta \dot{S}^{z}\right\rangle & =-\Gamma _{1}\left\langle
\delta S^{z}\right\rangle -\beta (S_{0}^{-}\left\langle \delta
f^{+}\right\rangle +f_{0}^{\ast }\left\langle \delta S^{-}\right\rangle
+S_{0}^{-\ast }\left\langle \delta f\right\rangle +f_{0}\left\langle \delta
S^{+}\right\rangle )  \notag \\
& -ig(S_{0}^{-\ast }\left\langle \delta a\right\rangle +a_{0}\left\langle
\delta S^{+}\right\rangle -S_{0}^{-}\left\langle \delta a^{+}\right\rangle
-a_{0}^{\ast }\left\langle \delta S^{-}\right\rangle )+i\Omega
_{pu}(\left\langle \delta S^{+}\right\rangle -\left\langle \delta
S^{-}\right\rangle )  \notag \\
& +(i\mu E_{pr}/\hbar )(S_{0}^{-\ast }e^{-i\delta t}-S_{0}^{-}e^{i\delta t})%
\text{,}
\end{align}%
\begin{eqnarray}
\left\langle \delta \dot{S}^{-}\right\rangle &=&-(i\Delta _{pu}+\Gamma
_{2})\left\langle \delta S^{-}\right\rangle +2\beta (S_{0}^{z}\left\langle
\delta f\right\rangle +f_{0}\left\langle \delta S^{z}\right\rangle
)+2ig(S_{0}^{z}\left\langle \delta a\right\rangle +a_{0}\left\langle \delta
S^{z}\right\rangle )  \notag \\
&&-2i\Omega _{pu}\left\langle \delta S^{z}\right\rangle -2i\mu
E_{pr}S_{0}^{z}e^{-i\delta t}/\hbar \text{,}
\end{eqnarray}%
\begin{equation}
\left\langle \delta \dot{f}\right\rangle =-(i\Delta _{M}+\kappa
_{M}/2))\left\langle \delta f\right\rangle +\beta \left\langle \delta
S^{-}\right\rangle \text{,}
\end{equation}%
\begin{equation}
\left\langle \delta \dot{a}\right\rangle =-(i\Delta _{n}+\kappa
/2))\left\langle \delta a\right\rangle -ig\left\langle \delta
S^{-}\right\rangle \text{,}
\end{equation}%
accordingly, we can obtain the corresponding steady-state equations and from
these equations we get: $\Gamma _{1}(w_{0}+1)\{(\Delta _{pu}^{2}+\Gamma
_{2}^{2})(\Delta _{M}^{2}+\kappa _{M}^{2}/4)(\Delta _{n}^{2}+\kappa
^{2}/4)+w_{0}^{2}[g^{4}(\Delta _{M}^{2}+\kappa _{M}^{2}/4)+\beta ^{2}(\Delta
_{n}^{2}+\kappa ^{2}/4)+g^{2}\beta ^{2}(2\Delta _{M}\Delta _{n}+\kappa
_{M}\kappa /4)]+w_{0}[g^{2}(\Delta _{M}^{2}+\kappa _{M}^{2}/4)(2\Delta
_{pu}\Delta _{M}-\Gamma _{2}\kappa _{M})+\beta ^{2}(2\Delta _{pu}\Delta
_{M}+\Gamma _{2}\kappa _{M})]\}-2\Omega _{pu}^{2}w_{0}[3\kappa g^{2}(\Delta
_{M}^{2}+\kappa _{M}^{2}/4)+\kappa _{M}\beta ^{2}(\Delta _{n}^{2}+\kappa
^{2}/4)]+4\Omega _{pu}^{2}w_{0}\Gamma _{2}(\Delta _{M}^{2}+\kappa
_{M}^{2}/4)(\Delta _{n}^{2}+\kappa ^{2}/4)=0$ which determines the
population inversion ($w_{0}=2S_{0}^{z}$) of the electron in CNT.

In order to solve Eqs. (7)-(10), we make the ansatz\cite{Boyd} $\left\langle
\delta S^{z}\right\rangle =S_{+}^{z}e^{-i\delta t}+S_{-}^{z}e^{i\delta t}$, $%
\left\langle \delta S^{-}\right\rangle =S_{+}e^{-i\delta t}+S_{-}e^{i\delta
t}$, $\left\langle \delta f\right\rangle =f_{+}e^{-i\delta
t}+f_{-}e^{i\delta t}$ and $\left\langle \delta a\right\rangle
=a_{+}e^{-i\delta t}+a_{-}e^{i\delta t}$. Substituting these equations into
Eqs. (7)-(10), and working to the lowest order in $E_{pr}$ but to all orders
in $E_{pu}$, we can obtain the linear susceptibility as $\chi
_{eff}^{(1)}(\omega _{pr})=\mu S_{+}(\omega _{pr})/E_{pr}=(\mu ^{2}/\hbar
)\chi ^{(1)}(\omega _{pr})$, where $\chi ^{(1)}(\omega _{pr})$ is given by%
\begin{equation}
\chi ^{(1)}(\omega _{pr})=\dfrac{(d_{4}^{\ast }-h_{2}d_{3}^{\ast
})d_{1}h_{3}-iw_{0}d_{4}^{\ast }}{d_{2}d_{4}^{\ast
}-h_{1}h_{3}d_{1}d_{3}^{\ast }}\Gamma _{2}\text{,}
\end{equation}%
where $f_{0}$, $S_{0}^{-}$ and $a_{0}$ can be derived from the steady-state
equations and $b_{1}=\beta /(i\Delta _{M}+\kappa _{M}/2-i\delta )$, $%
b_{2}=\beta /(i\Delta _{M}+\kappa _{M}/2+i\delta )$, $b_{3}=-ig/(i\Delta
_{n}+\kappa /2-i\delta )$, $b_{4}=-ig/(i\Delta +\kappa /2+i\delta )$, $%
h_{1}=[-ig(S_{0}^{-\ast }b_{3}-a_{0}^{\ast })-\beta (f_{0}^{\ast
}+S_{0}^{-\ast }b_{1})-i\Omega _{pu}]/(\Gamma _{1}-i\delta )$, $%
h_{2}=[-ig(a_{0}-S_{0}^{-}b_{4}^{\ast })-\beta (S_{0}^{-}b_{2}^{\ast
}+f_{0})+i\Omega _{pu}]/(\Gamma _{1}-i\delta )$, $h_{3}=iS_{0}^{-\ast
}/(\Gamma _{1}-i\delta )$, $d_{1}=2(iga_{0}-i\Omega _{pu}+\beta f_{0})$, $%
d_{2}=i\Delta _{pu}+\Gamma _{2}-i\delta -(igb_{3}+\beta
b_{1})w_{0}-d_{1}h_{1}$, $d_{3}=2(iga_{0}-i\Omega _{pu}+\beta f_{0})$, $%
d_{4}=i\Delta _{pu}+\Gamma _{2}+i\delta -(igb_{4}+\beta
b_{2})w_{0}-d_{3}h_{4}$ (where $O^{\ast }$ indicates the conjugate of $O$).
The quantum Langevin equations of the normal electrons coupled to CNT have
the same form as Majorana fermions, therefore, we omit its derivation and
only give the results in the following section.

\section{RESULTS AND DISCUSSIONS}

The electron with spin trapped inside the center of CNT couples to the
nearby Majorana fermion in InSb nanowire in the simultaneous presence of a
strong pump beam and a weak probe beam as shown in Fig.1. For illustration
of the numerical results, here we use the realistic experimental parameters%
\cite{Kuemmeth:nat08,Steele:sci09,Struck:prl12}: $\kappa =0.05MHz$, $\Gamma
_{1}=0.1MHz$, $\Gamma _{2}=\Gamma _{1}/2$ and the coupling strength between
the electron spin and the nanotube vibrational mode $g=2\pi \times 0.56MHz$.
For Majorana fermion, there are no experimental values for the lifetime of
the Majorana fermion and the coupling strength between the electron spin and
Majorana bound states as far as we know. However, according to a few reports
\cite{Mao:prl12,Deng,Mourik:sci12,Rokhinson,Das}, we assume that the order
of magnitudes of these two parameters are in the range of megahertz, while
in our letter, we consider the coupling strength $\beta =0.2MHz$ and the
lifetime of the Majorana fermion $\kappa _{M}=0.1MHz$. When the electron
spin in CNT couples to the nearby MBS and produces the coupled states $%
\left\vert ex,n_{M}\right\rangle $ and $\left\vert ex,n_{M}+1\right\rangle $
($n_{M}$ denotes the number states of the Majorana bound states) as shown in
the inset of Fig. 1.

When radiating a strong pump field and a weak probe field on this system, we
can detect the existence of a Majorana fermion via the coupling between
Majorana bound states and the electron spin in CNT from the probe absorption
spectrum. Fig. 2(a) shows the absorption spectrum of the probe field
(i.e.,the imaginary part of the dimensionless susceptibility $Im\chi ^{(1)}$%
) as a function of the probe-spin detuning $\Delta _{pr}$ ($\Delta
_{pr}=\omega _{pr}-\omega _{S}$) with several different coupling strengths $%
\beta $. The black solid curve is the result when no Majorana bound states
exists in the nanowire. As Majorana bound states appears at the ends of the
nanowire, the electron spin in CNT will couple to the right Majorana bound
states, which induces the upper level of the state $\left\vert
ex\right\rangle $ to split into $\left\vert ex,n_{M}\right\rangle $ and $%
\left\vert ex,n_{M}+1\right\rangle $. The two peaks presented in Fig. 2(a)
with a given coupling strength $\beta $ indicate the spin-MBS interaction.
As shown in the low two insets of Fig. 2(a), the left peak signifies the
transition from $\left\vert g\right\rangle $ to $\left\vert
ex,n_{M}\right\rangle $ while the right peak is due to the transition of $%
\left\vert g\right\rangle $ to $\left\vert ex,n_{M}+1\right\rangle $. To
determine this signature is the true Majorana bound states that appear in
the nanowire, rather than the normal electrons that couple with the QD, we
give the numerical results of the normal electrons in the nanowire that
couple with the QD as shown in the upper right inset of Fig.2(a). In order
to compare with the Majorana signature, the parameters are chosen the same
as Majorana fermion's parameters. We find that there is no signal in the
probe absorption spectrum (see the red solid line in the inset of Fig.2(a))
which means that the splitting of the probe absorption spectrum is the true
signature of MBS. Therefore, our result here reveals that the splitting in
the probe absorption is a real signature of MBS, and the stronger coupling
strength introduces the wider and deeper dip. Furthermore, the distance of
the splitting becomes larger and larger with increasing the coupling
strength $\beta $, which obviously reveals the spin-MBSs coupling. From Fig.
2(b) we find that the distance of the splitting is twice times larger than
the spin-MBSs coupling strength, which provides a concise way to measure the
spin-MBS coupling strength in this coupled system. In Fig. 2(b), we give a
linear relationship between the distance of the peak splitting and the
spin-MBS coupling strength. Therefore, the coupling strength can be obtained
immediately by directly measuring the distance of two peaks in the probe
absorption spectrum.

Fig. 3 demonstrates the behavior of the lifetime of the Majorana fermion
when the pump field is resonant with the frequency of the electron spin in
CNT. In Fig. 3, we plot the relationship between the transmission and the
lifetime of the Majorana fermion. It shows that the transmission of the
probe field decreases with the increase of the decay rate of the Majorana
fermion. This figure provides us a simple method to measure the decay rate
of the Majorana fermion. Through this method, we can determine the decay
rate of the Majorana fermion and further investigate its relation with the
environment via the transmission spectrum of the probe field.

In order to demonstrate the function of the vibration in CNT, we plot Fig.
4. Fig. 4 presents the absorption spectrum of the probe field as a function
of the probe detuning $\Delta _{pr}$ with different coupling strengths $g$.
In Fig. 4, three couplings strength $g=0$, $0.2$, $0.56MHz$ are considered. $%
g=0$ means that there is no spin-phonon coupling in the suspended CNT, the
two sharp peaks in the absorption spectrum still appear. However, once the
spin-phonon coupling turns on and increases the coupling to the realistic
experimental value $g=0.56MHz$ \cite{Kuemmeth:nat08}, the two peaks of the
absorption spectrum becomes more significant than that without the
spin-phonon coupling in CNT, which means that the vibration of CNT (i.e. the
phonon cavity) enhances the probe spectrum and makes the Majorana fermion
more sensitive to be detected. This phonon cavity enhanced effect is
analogous to the optical cavity enhanced effect in quantum optics\cite%
{Stapfner}.

\section{CONCLUSIONS}

In conclusion, a novel scheme which consists of an electron spin trapped on
the suspended carbon nanotube and a tunnel-coupled semiconductor nanowire
for demonstrating the existence of a Majorana fermion is proposed according
to spin-based optomechanics. Based on this scheme, we propose a simple
method to determine the interaction constant of the spin-Majorana bound
states and the decay rate of the Majorana fermion via the probe absorption
spectrum. Due to the phonon cavity enhanced effect, the Majorana fermion
becomes more and more easier to be detected in this coupled system. Our
scheme proposed here may provide us a potential application in all-optical
controlled Majorana fermion based quantum computer and quantum information
processing.

\section{ACKNOWLEDGMENTS}

The authors gratefully acknowledge support from the National Natural Science
Foundation of China (No.10974133 and No.11274230), the National Ministry of
Education Program for PhD and the National Basic Research Program of the
Ministry of Science and Technology of China.

\centerline{\large{\bf Figure Captions}}

FIG.1 Schematic setup for detecting a Majorana fermion through an electron
spin trapped on the suspended CNT coupled to a semiconductor nanowire
contacting with a s-wave superconductor in the simultaneous presence of a
strong pump field and a weak probe field. The inset is an energy-level
diagram of the electron spin coupled to Majorana fermion in semiconductor
nanowire and the vibrational motion in CNT.

FIG.2 (a) The probe absorption spectrum as a function of detuning $\Delta
_{pr}$ with several different spin-MBS coupling strengths. The black solid
line is the result for $\beta =0$MHz and $\lambda =0$MHz. The upper right
inset is the normal electrons in the nanowire that couple with the CNT at
the coupling strength $\lambda =0.3$MHz and the spin-MBS coupling strength $%
\beta =0.3$MHz. The low two insets represents the energy level transitions
of the left peak and right peak presented in the absorption spectrum. (b)
The linear relationship between the distance of peak splitting and the
coupling strength of spin-MBS. The parameters used are $\Gamma _{1}=0.1$MHz,
$\Gamma _{2}=\Gamma _{1}/2$, $\kappa _{M}=0.1$MHz, $\kappa =0.05$MHz, $g=0$%
MHz, $\Delta _{M}=-0.1$MHz, $\Omega _{pu}^{2}=7\times 10^{-4}($MHz$)^{2}$
and $\Delta _{pu}=\Delta _{n}=0$.

FIG.3 The transmission of a probe beam as a function of the lifetime of
Majorana fermion. The parameters used are $\Gamma _{1}=0.1$MHz, $\Gamma
_{2}=\Gamma _{1}/2$, $\beta =0.3$MHz, $\kappa =0.05$MHz, $g=0$MHz, $\Delta
_{M}=-0.1$MHz, $\Omega _{pu}^{2}=7\times 10^{-4}($MHz$)^{2}$ and $\Delta
_{pu}=\Delta _{n}=0$.

FIG.4 The probe absorption spectrum as a function of detuning $\Delta _{pr}$
with three different spin-phonon coupling strengths $g=0$, $0.2$, $0.56MHz$.
The parameters used are the same as in FIG.3.
\begin{figure}[tbp]
\centering
\includegraphics[width=12cm]{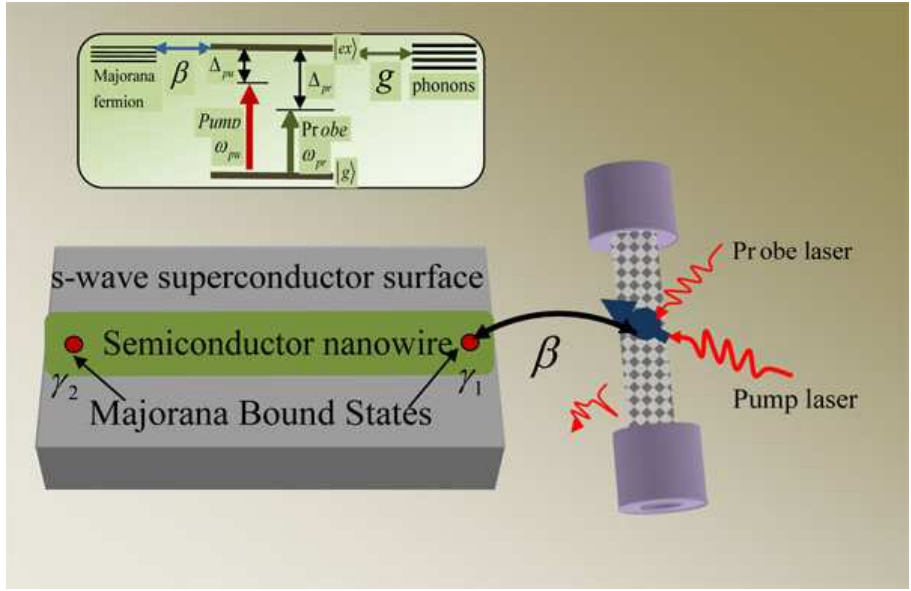}
\caption{Schematic setup for detecting a Majorana fermion through an
electron spin trapped on the suspended CNT coupled to a semiconductor
nanowire contacting with a s-wave superconductor in the simultaneous
presence of a strong pump field and a weak probe field. The inset is an
energy-level diagram of the electron spin coupled to Majorana fermion in
semiconductor nanowire and the vibrational motion in CNT. }
\end{figure}

\begin{figure}[tbp]
\centering
\includegraphics[width=12cm]{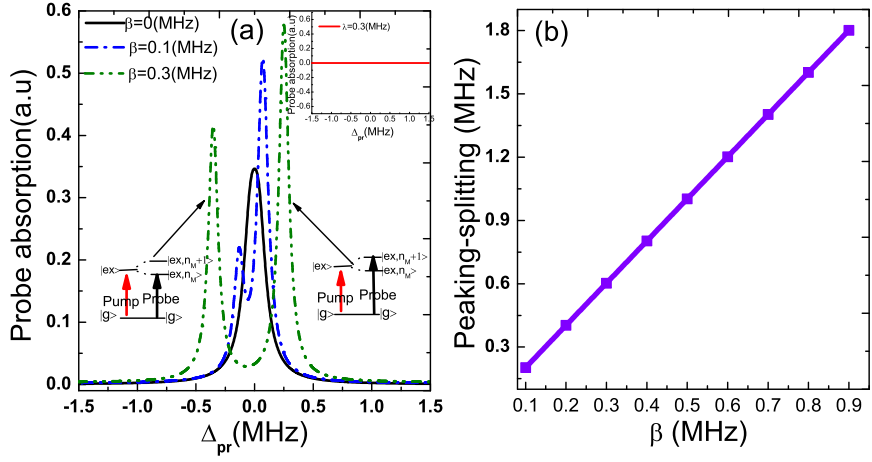}
\caption{(a) The probe absorption spectrum as a function of detuning $\Delta
_{pr}$ with several different spin-MBS coupling strengths. The black solid
line is the result for $\protect\beta =0$MHz and $\protect\lambda =0$MHz.
The upper right inset is the normal electrons in the nanowire that couple
with the CNT at the coupling strength $\protect\lambda =0.3$MHz and the
spin-MBS coupling strength $\protect\beta =0.3$MHz. The low two insets
represents the energy level transitions of the left peak and right peak
presented in the absorption spectrum. (b) The linear relationship between
the distance of peak splitting and the coupling strength of spin-MBS. The
parameters used are $\Gamma _{1}=0.1$MHz, $\Gamma _{2}=\Gamma _{1}/2$, $%
\protect\kappa _{M}=0.1$MHz, $\protect\kappa =0.05$MHz, $g=0$MHz, $\Delta
_{M}=-0.1$MHz, $\Omega _{pu}^{2}=7\times 10^{-4}($MHz$)^{2}$ and $\Delta
_{pu}=\Delta _{n}=0$. }
\end{figure}

\begin{figure}[tbp]
\centering
\includegraphics[width=12cm]{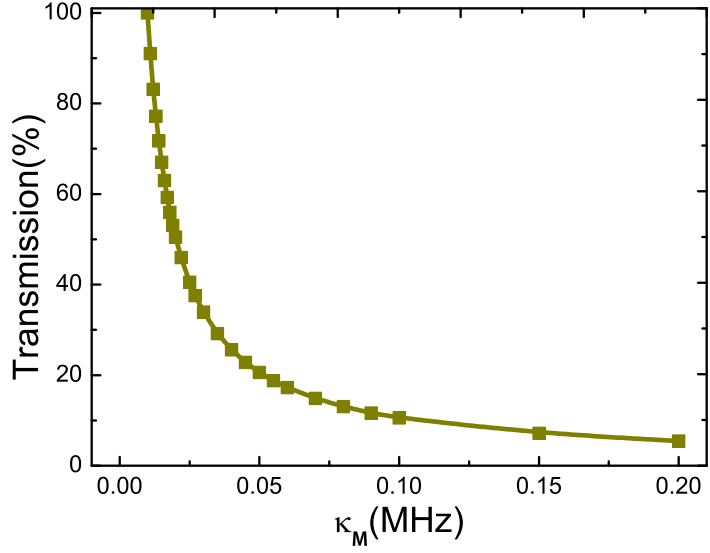}
\caption{The transmission of a probe beam as a function of the lifetime of
Majorana fermion. The parameters used are $\Gamma _{1}=0.1$MHz, $\Gamma
_{2}=\Gamma _{1}/2$, $\protect\beta =0.3$MHz, $\protect\kappa =0.05$MHz, $%
g=0 $MHz, $\Delta _{M}=-0.1$MHz, $\Omega _{pu}^{2}=7\times 10^{-4}($MHz$%
)^{2} $ and $\Delta _{pu}=\Delta _{n}=0$.}
\end{figure}

\begin{figure}[tbp]
\centering
\includegraphics[width=12cm]{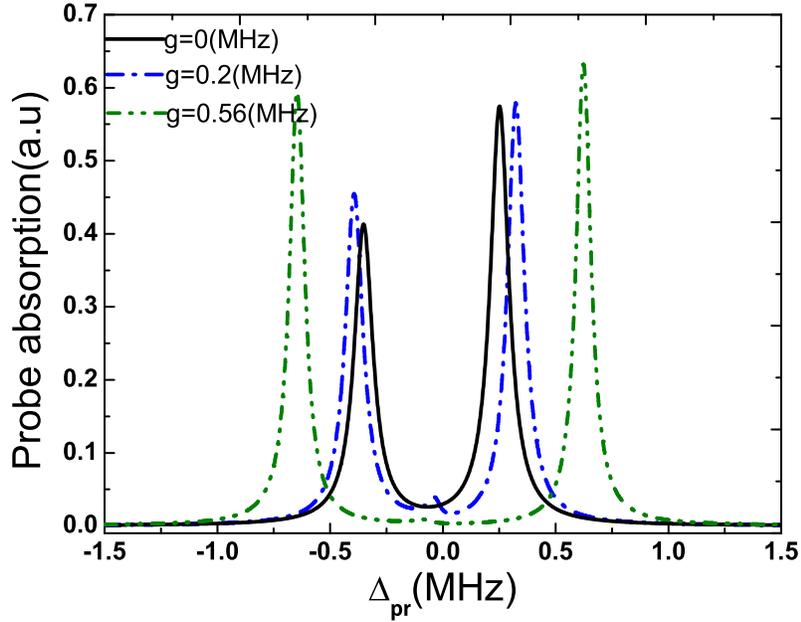}
\caption{The probe absorption spectrum as a function of detuning $\Delta
_{pr}$ with three different spin-phonon coupling strengths $g=0$, $0.2$, $%
0.56MHz$. The parameters used are the same as in FIG.3. }
\end{figure}

\end{document}